# Conceptual Preconditions of Overcoming of Relativistic Intentions in Modern Philosophy of Science

*Sergey B. Kulikov*

*Tomsk State Pedagogical University*

E-mail: kulikovsb@tspu.edu.ru

ABSTRACT: The paper defends the thesis that it's possible to maintain some conceptual preconditions of overcoming of relativistic intentions in modern philosophy of science ("there are no any general foundations in philosophy of science"). We found two general foundations in philosophy of science as a minimum. From the first side it's realistic to reveal on the base of special understanding of time the meaning of time not only in natural thought (especially in theory of gravity) but also in humanitarian knowledge. That's why philosophy of science has independent position in epistemology and ontology corresponding to interpretation of time as a general category of scientific thinking. The nature of time has internally inconsistent (paradoxical) character. Time is phenomenon which existing and not existing at the same time. This phenomenon is identified with imaginary movement and also ideal (formal) process of formation of the nature. The general understanding of time is connected with its "mathematical" meaning as calculable formal regulation of language practice and also the universal organization rules of quantitative parameters of intelligence of natural (physical) processes. From the second side we can say that exist an actual branch of philosophy of science. It exists on the basis of disclosure of aprioristic limits of consciousness of its cultural and historical development. There is possible a special interpretation of time. In that context time is the connection of an action of the cultural phenomenon or its "energy" with some kind of "weight", the historical importance of a separate limit of consciousness through analog of "distance" as intensity of cultural and historical space (or "oppositional nature of interaction of mental intentions").

# Introduction: Main Approaches to the Foundations of Philosophy of Science

Relevance of a question of conceptual prerequisites of overcoming of relativism in modern philosophy of science is caused by an essential theoretical divergence between the intensions of classical thought connected with idea of the organization of integrity of scientific knowledge, and also modern approaches to understanding of the idea of the organization of science. Within classics I. Kant allocating metaphysical fundamentals of natural sciences used distinction of ways of representation of the nature in formal and their material meaning [1, p. 987]. It assumed that "... any doctrine if it system, i.e. the certain set of knowledge ordered in compliance with the principles, is called as science. ... In the true sense it is possible to call science only that, which reliability apodictic. ... The science about the nature acquires the right to be called that only from pure part" [1, p. 988-989]. Thus, from the classical point of view, the knowledge comprising system of aprioristic laws admits scientific or able to be developed in that quality. And to this ideal most fully there corresponds mathematics, in particular Euclidean geometry and arithmetic [1, p. 990].

Idea of specific characteristics of aprioristic knowledge acts as the basis of a classical position. According to I. Kant, aprioristic knowledge receive proceeding from possible ("pure") representations of a cognizable subject independent of experience [1, p. 991]. Such knowledge is caused by the requirement of a special form of contemplation. This form is based on designing of concepts corresponding to the deductive organization of bases of mathematics. Therefore though the pure philosophy about the "highest" or "formal" nature (i.e. the principles of things) also is possible without mathematics, any doctrine about concrete (financially certain) nature from mathematics continuously and "...will contain science in the true sense only in that measure in what the mathematics" [1, p. 991]. Thus, the carrier of formal (or transcendental) truth is the "pure philosophy" or metaphysics, knowledge from "only one concepts" [1, p. 989]; as the carrier of material (or "concrete") truth the mathematics acts. Many other classical authors (R. Descartes, B. Spinoza, G. Leibniz, G. Hegel, etc.) adhered as a whole to similar views for a mathematics role in structure of general scientific knowledge.

In work [2] was proved two theses:

(1) in modern philosophy of science there is a radical reconsideration of classical positions in favor of statement under a question of meaning of deductive knowledge as itself, and to it – unities of bases of science (i.e. their relativization);

(2) the relativism in modern philosophy of science directly follows from provisions of E. Husserl directed on criticism of the base of physical and mathematical natural sciences.

Relativistic views got even into deduction kingdom – formal logic in the form of idea of relative randomness of creation of axiomatic theories and application of methods of the logical analysis [3, p. 5; 4, p. 2-3]. That's why they say that "there are no any general foundations in philosophy of science".

## Has Philosophy of Science an Independent Position in the Scopes of Epistemological and Ontological Studies?

Disclosure of conceptual prerequisites of overcoming of a modern relativism becomes possible in the conditions of interpretation in work [5] philosophy of science as rather independent branch of fundamental knowledge, the unity of which subject domain isn't deduced and isn't synthesized, but is set as special opposition of research culture as a limit of her conscious consciousness. The contents of these prerequisites are directly connected with two moments:

(1) allocation of a role of philosophy of science in ontological and epistemological studies;

(2) disclosure of basis of development of philosophy of the science serving as the base of the possibility of understanding of the valid processes.

In this regard in modern thought keeps relevance an important theoretical contradiction. On the one hand, K. Popper agrees that "natural sciences and natural philosophy are faced by a grandiose task – to create a coherent and clear picture of the Universe" [6, p. 13]. On another hand, the solution of this task is essentially complicated within modern concepts [5, p. 11-20]. We believe that the exit here can be found in the scope of analysis of conceptualization bases of time in classical and modern philosophy.

The argument of our thesis means, first, the statement that time is among fundamental categories of natural sciences. It is caused by an orientation of natural-science thought on allocation of structure and regularities of course of natural processes. Secondly, as M. Heidegger notices, "the science in general can be defined as set of justifying interrelation of true provisions. This definition both isn't full, and doesn't catch science in its sense. Sciences as images of behavior of the person have a way of life of this real (person). Real we seize this terminologically as presence. ... As proof of that – and as – temporariness constitutes presence life, it was shown: historicity as the being's device of an existence is "in a basis" temporariness" [7, p. 27, 451]. From all this follows that the category of time in natural sciences as an element of general scientific knowledge sends not only to area of natural processes, but also their sense concerning a life perspective in general and questions of life of the person in particular.

Therefore, we also believe that the role (independent position) of philosophy of science in ontological and epistemological studies can be established during the analysis of bases of conceptualization of time in classical and modern philosophy. One of the ways of justification of philosophy of science opens through detection of its specifics in the general framework of ontological studies as problematic relation of reality/people.

## Special Meaning of Time as a General Foundation of Philosophy of Science

In work [8] the following was revealed:

1. The nature of time has internally inconsistent (paradoxical) character and consists in unity of its existence and not existence.

2. Time is identified with imaginary movement in the Greek classics, and also ideal (formal) process of formation of the nature in the German classical thought.

3. The general understanding of time is connected with its "mathematic meaning", i.e. formality and a universalism as regulation of language practice, according to Aristotle, and also the organization of quantitative parameters of intelligence of natural processes, according to G. Hegel.

Further researches showed that, in particular, J.-P. Sartre during the analysis of ideas of R. Descartes, I. Kant, G. Hegel, E. Husserl, A. Bergson and other thinkers comes to a conclusion that from the philosophical point of view on the nature and structure of reality it is impossible to reach the rational solution of the task, what concrete appearance the science basis has: phenomenology of consciousness or consciousness and life opposition. The solution of such task can be executed pragmatically on the basis of a metaphysical postulate on special type of informative "benefit". In this regard time only in opportunity belongs to the nature as, "... as in Einstein's physics where find it possible to speak about a cognizable event as about having spatial measurements and one temporary measurement and defining the place in space time" [9, p. 624]. This conclusion assumes that is lawful both non-classical approach and traditional approach. Non-classical approach includes representation that the world and its attributes are a phenomenon of human consciousness. Framework of traditional views includes opposition between the subject and object (consciousness and life distinction) and space and time opposition, for example, in the form of a point and number (i.e. locations of object, and also a measure of its formation [10, p. 51-59]) is meant.

All of this nowadays essentially complicates philosophical justification of natural sciences.

We take the principled stand which fundamental prerequisites are in this relation: 1) idea of discretization of cultural and historical process, and also 2) virtual nature of interaction of conscious limits of consciousness of philosophical and scientific knowledge (elements of limitation are found in real history and are in the oppositional relation to last and (or) hypothetically possible future moments) [5, p. 11-44]. Our position sends to a complex of philosophical provisions, but also can be formulated in natural-science categories of a time point as *t*, energy as *E*, *m* as a body weight, and also *r* as distances (enter the classical theory of gravitation (Newton) and relativistic mechanics (Einstein)). There is possible a special interpretation in which context time connects action of the phenomenon or its "energy" with some kind of "weight", the historical importance of a separate limit of consciousness through distance analog as intensity of cultural and historical space, i.e. oppositional nature of interaction of intentions. The complex of philosophical ideas which allow finding a way of rational justification of natural sciences comes to light, coordinating modern level of philosophical understanding of time and life and elements of natural-science knowledge.

## Conclusion

Thus, conceptual preconditions of overcoming of relativistic intentions in modern philosophy of science are:

(1) allocation of a role of philosophy of science in ontological and epistemological studies which opens in the light of formation of special understanding of time revealing meaning of this category not only in natural-science, but also in humanitarian thought;

(2) branch division of philosophy of science which could be made on the basis of disclosure of aprioristic limits of conscious consciousness of its cultural historical development.

All of this opens prospect of cognition of natural and cultural and historical processes on the basis of a community of rather independent complexes of natural and humanitarian disciplines, their active interaction and mutual enrichment.